\documentclass[10pt]{article}
\usepackage{fullpage}
\usepackage{amsmath}
\usepackage{amssymb}
\usepackage{amsfonts}
\usepackage[dvips]{epsfig}
\usepackage{color}

\def\##1{\underline{#1}}
\def\=#1{\underline{\underline{#1}}}

\def\+#1{\underline{\bf #1}}
\def\*#1{\underline{\underline{\bf #1}}}

\def\r#1{(\ref{#1})}
\def\l#1{\label{#1}}
\def\c#1{\cite{#1}}

\def\le{\left(}
\def\ri{\right)}

\def\lec{\left\{}
\def\ric{\right\}}

\def\.{\mbox{ \tiny{$^\bullet$} }}

\def\eps{\epsilon}

\begin{document}

\begin{center}

{\bf {\LARGE Bruggeman formalism vs. `Bruggeman formalism':
 Particulate composite materials comprising oriented ellipsoidal particles}}

\vspace{10mm} \large

 Tom G. Mackay\footnote{E--mail: T.Mackay@ed.ac.uk}\\
{\em School of Mathematics and
   Maxwell Institute for Mathematical Sciences\\
University of Edinburgh, Edinburgh EH9 3JZ, UK}\\
and\\
 {\em NanoMM~---~Nanoengineered Metamaterials Group\\ Department of Engineering Science and Mechanics\\
Pennsylvania State University, University Park, PA 16802--6812,
USA}\\
 \vspace{3mm}
 Akhlesh  Lakhtakia\footnote{E--mail: akhlesh@psu.edu}\\
 {\em NanoMM~---~Nanoengineered Metamaterials Group\\ Department of Engineering Science and Mechanics\\
Pennsylvania State University, University Park, PA 16802, USA}

\end{center}

\vspace{15mm}

\begin{abstract}
Two different formalisms for the homogenization of composite
materials containing oriented ellipsoidal particles of isotropic
dielectric materials are being named after Bruggeman. Numerical
studies reveal clear differences between the two formalisms which
may  be exacerbated:
 (i) if the component particles become more aspherical, (ii)
at mid-range values of the volume fractions, and (iii) if the
homogenized component material is dissipative. The correct Bruggeman
formalism uses the correct polarizability density dyadics of the
component particles, but the other formalism does not.
\end{abstract}

\noindent {\bf keywords:} Bruggeman homogenization formalism,
homogenized composite materials, ellipsoidal particles

\vspace{5mm}

%\section{Introduction}

The Bruggeman formalism provides a well-established technique for
estimating  the effective constitutive parameters of homogenized
composite materials (HCMs) \c{Ward,L96,M11}. The scope of its applicability   is not restricted to
dilute composite materials and it is easy to implement numerically, both of which contribute
to its enduring popularity.

The Bruggeman formalism was originally devised for
isotropic dielectric HCMs, comprising two (or more) isotropic dielectric
component materials   distributed randomly as
electrically small spherical particles \c{Br}. Generalizations of
the Bruggeman formalism which accommodate anisotropic and
bianisotropic HCMs have been developed \c{WLM}.
 A rigorous basis
for the Bruggeman formalism---for isotropic dielectric \c{TK81},
anisotropic dielectric \c{Genchev,Z94}, and
bianisotropic \c{MLW00,MLW01} HCMs---is
provided by the strong-property-fluctuation theory, whose lowest-order formulation is
 the Bruggeman formalism.

Our focus in this letter is on HCMs arising from two isotropic
dielectric component materials, labeled $a$ and $b$. Their relative
permittivities are $\eps^a$ and $\eps^b$, while their volume
fractions are $f_a $ and $f_b \equiv 1- f_a$. Both component
materials are assumed to be randomly distributed as electrically
small ellipsoidal particles. For simplicity, all component particles
have the same shape and orientation. The surface of each ellipsoid,
relative to its centroid, may be represented by the vector
\begin{equation}
\#r_{\,e} (\theta, \phi) = \eta \, \=U \. \hat{\#r} (\theta, \phi),
\end{equation}
with $ \hat{\#r} $ being the radial unit vector from the ellipsoid's
centroid, specified by the spherical polar coordinates $\theta$ and
$\phi$. The linear dimensions of each ellipsoid, as determined by
the parameter $\eta$, are  assumed to be small relative to the
electromagnetic wavelength(s). Let us choose our coordinate system
to be such that the Cartesian axes are aligned with the principal
axes of the ellipsoids. Then
 the ellipsoidal shape is captured by the dyadic
\begin{equation}
 \=U =  U_x \, \hat{\#x} \, \hat{\#x} + U_y \, \hat{\#y} \, \hat{\#y} + U_z \, \hat{\#z} \, \hat{\#z} ,
\end{equation}
wherein the shape parameters $U_{x,y,z} > 0$ and $\lec  \hat{\#x},
\hat{\#y},  \hat{\#z} \ric$ are unit vectors aligned with the
Cartesian axes.

The ellipsoidal shape of the component particles results in the
corresponding HCM being an orthorhombic biaxial dielectric material.
That is, the Bruggeman estimate of the HCM relative permittivity
dyadic has the form
\begin{equation}
 \=\eps^{Br1} =  \eps^{Br1}_x \, \hat{\#x} \, \hat{\#x} + \eps^{Br1}_y \, \hat{\#y} \, \hat{\#y} + \eps^{Br1}_z \, \hat{\#z} \, \hat{\#z}
 .
\end{equation}
The relative permittivity parameters $\eps^{Br1}_{x,y,z}$ are
 given implicitly by the three coupled equations \c{ML_Prog_Opt}
\begin{eqnarray}
&& \frac{\eps^a - \eps^{Br1}_\ell }{1 + D_\ell \le \eps^a -
\eps^{Br1}_\ell \ri} f_a + \frac{\eps^b - \eps^{Br1}_\ell }{1 +
D_\ell \le \eps^b - \eps^{Br1}_\ell \ri} f_b = 0 \,, \qquad \le \ell
\in \lec x, y, z \ric \ri. \l{Br1}
\end{eqnarray}
Herein $D_\ell$ are components of the depolarization dyadic
\begin{equation}
\=D = D_x \hat{\#x} \, \hat{\#x} + D_y \hat{\#y} \, \hat{\#y} + D_z
\hat{\#z} \, \hat{\#z},
\end{equation}
where the double integrals\c{WLM}
\begin{equation}
\left.
\begin{array}{l}
D_x = \displaystyle{\frac{1}{ 4 \pi} \, \int^{2 \pi}_0 \, d \phi \,
\int^\pi_0 \, d \theta \,  \,  \frac{\sin^3 \theta \, \cos^2
\phi}{U^2_x \, \rho}} \vspace{8pt} \\
D_y = \displaystyle{\frac{1}{ 4 \pi} \, \int^{2 \pi}_0 \, d \phi \,
\int^\pi_0 \, d \theta \,  \,  \frac{\sin^3 \theta \, \sin^2
\phi}{U^2_y \, \rho}} \vspace{8pt} \\
D_z = \displaystyle{\frac{1}{ 4 \pi} \, \int^{2 \pi}_0 \, d \phi \,
\int^\pi_0 \, d \theta \,  \,  \frac{\sin \theta \, \cos^2
\theta}{U^2_z \, \rho}}
\end{array}
\right\},  \l{depol_x}
\end{equation}
involve the scalar parameter
\begin{equation}
\rho = \frac{\sin^2 \theta \, \cos^2 \phi}{U_x^2}\, \eps^{Br1}_x  +
\frac{\sin^2 \theta \, \sin^2 \phi}{U_y^2}\, \eps^{Br1}_y +
\frac{\cos^2 \theta}{U_z^2}\, \eps^{Br1}_z.
\label{rho}
\end{equation}
The coupled nature of the three
eqns. \r{Br1} means that numerical methods are generally needed to
extract the relative permittivity parameters $\eps^{Br1}_{x,y,z}$
from them.

An alternative  formalism for the homogenization of the same composite material
as in the foregoing paragraph is also referred to as
 the Bruggeman formalism
\c{Smith,Granqvist,Schubert1,Schubert2}.  Let us write the estimate  of the HCM's relative permittivity dyadic
 provided by this alternative formalism as
\begin{equation}
 \=\eps^{Br2} =  \eps^{Br2}_x \, \hat{\#x} \, \hat{\#x} + \eps^{Br2}_y \, \hat{\#y} \, \hat{\#y} + \eps^{Br2}_z \, \hat{\#z} \,
 \hat{\#z}.
\end{equation}
The relative permittivity parameters $\eps^{Br2}_{x,y,z}$ are
 given  by the three equations
\begin{eqnarray}
&& \frac{\eps^a - \eps^{Br2}_\ell }{\eps^{Br2}_\ell + L_\ell \le
\eps^a - \eps^{Br2}_\ell \ri} f_a + \frac{\eps^b - \eps^{Br2}_\ell
}{\eps^{Br2}_\ell + L_\ell \le \eps^b - \eps^{Br2}_\ell \ri} f_b = 0
\,, \qquad \le \ell \in \lec x, y, z \ric \ri, \l{Br2}
\end{eqnarray}
wherein the depolarization factors \c{PvS}
\begin{equation}
L_\ell = \frac{U_x U_y U_z}{2} \int^\infty_0 \; ds \frac{1}{\le s +
U^2_\ell \ri \sqrt{\le s + U^2_x \ri \le s + U^2_y \ri \le s + U^2_z
\ri}}\,, \qquad \le \ell \in \lec x, y, z \ric \ri \l{L}
\end{equation}
 are components of the depolarization dyadic
\begin{equation}
\=L = L_x \hat{\#x} \, \hat{\#x} + L_y \hat{\#y} \, \hat{\#y} + L_z
\hat{\#z} \, \hat{\#z}.
\end{equation}
Each of the three eqns.~\r{Br2}  is a quadratic equation in
$\eps^{Br2}_\ell$ whose solution may be explicitly expressed as
\begin{equation}
\eps^{Br2}_\ell = \frac{- \beta \pm \sqrt{\beta^2 - 4 \alpha
\gamma}}{2\alpha}\,, \qquad \le \ell \in \lec x, y, z \ric \ri,
\l{eBr2_sol}
\end{equation}
with $\alpha = L_\ell - 1$,
$\beta =  \eps^a \le f_a - L_\ell\ri + \eps^b \le f_b - L_\ell \ri$, and
$\gamma = L_\ell \eps^a \eps^ b$.
The sign of the square root term in the solution \r{eBr2_sol} may be
determined by appealing to the anisotropic dielectric generalization
of the Hashin--Shtrikman bounds \c{H-S_bounds}, for example.

\begin{figure}[!ht]
\centering
\includegraphics[width=2.6in]{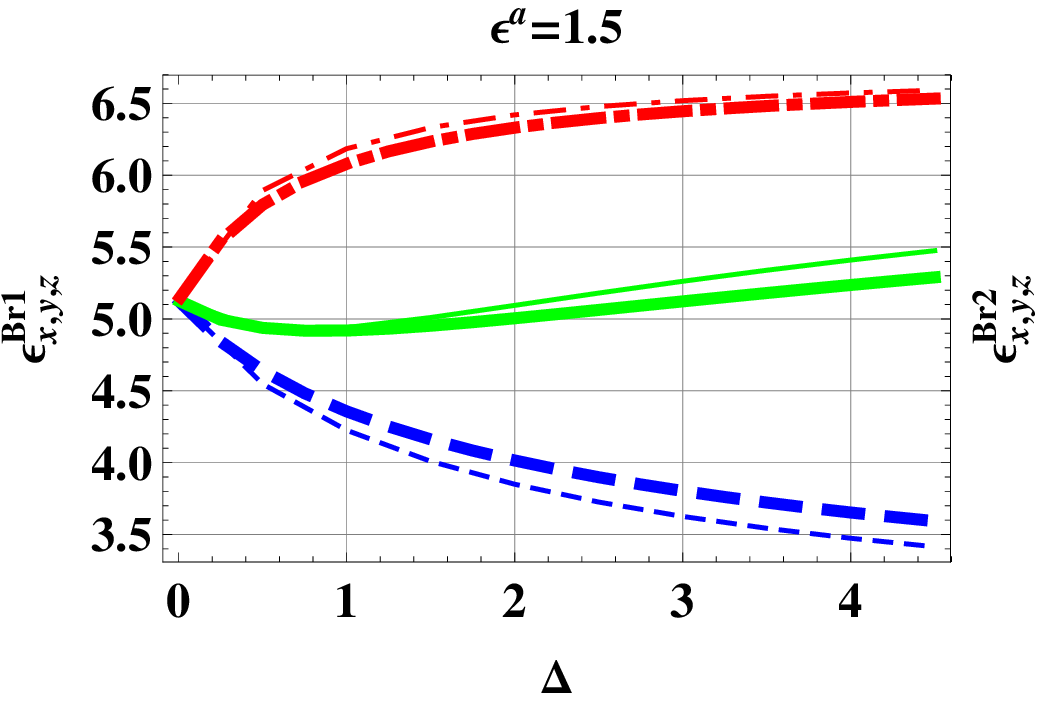} \hfill
\includegraphics[width=2.6in]{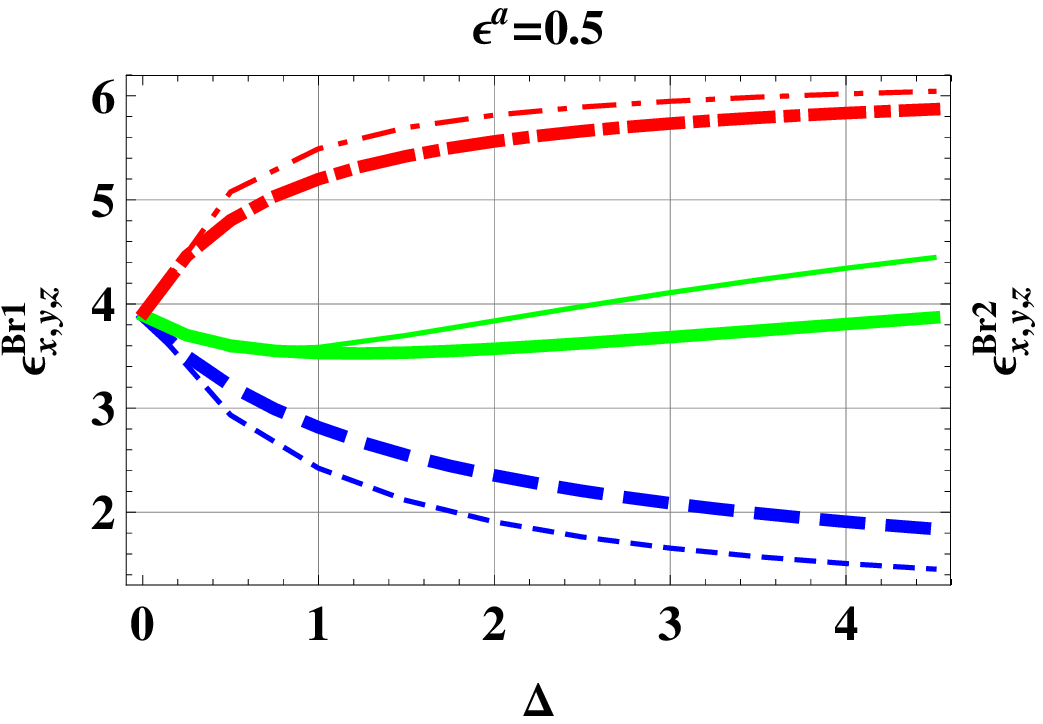}
 \caption{\l{fig1}
The estimates $\eps^{Br1,2}_x$ (blue, dashed curves),
$\eps^{Br1,2}_y$ (green, solid curves), and $\eps^{Br1,2}_z$ (red,
broken dashed curves) plotted versus the asphericity parameter
$\Delta \in \le 0, 4.5 \ri$. The $\eps^{Br1}_{x,y,z}$ estimates are
represented by thick curves and the $\eps^{Br2}_{x,y,z}$ estimates
are represented by thin curves. The ellipsoidal shapes of the
component material particles  are described by shape parameters $U_x
= 1$, $U_y = 1 + (\Delta/3)$, and $U_z = 2 \Delta$. The relative
permittivities of the component materials are $\eps^a \in \lec 0.5,
1.5 \ric$ and $\eps^b = 12$; and the volume fraction $f_a = 0.5$.}
\end{figure}

\begin{figure}[!h]
\centering
\includegraphics[width=2.6in]{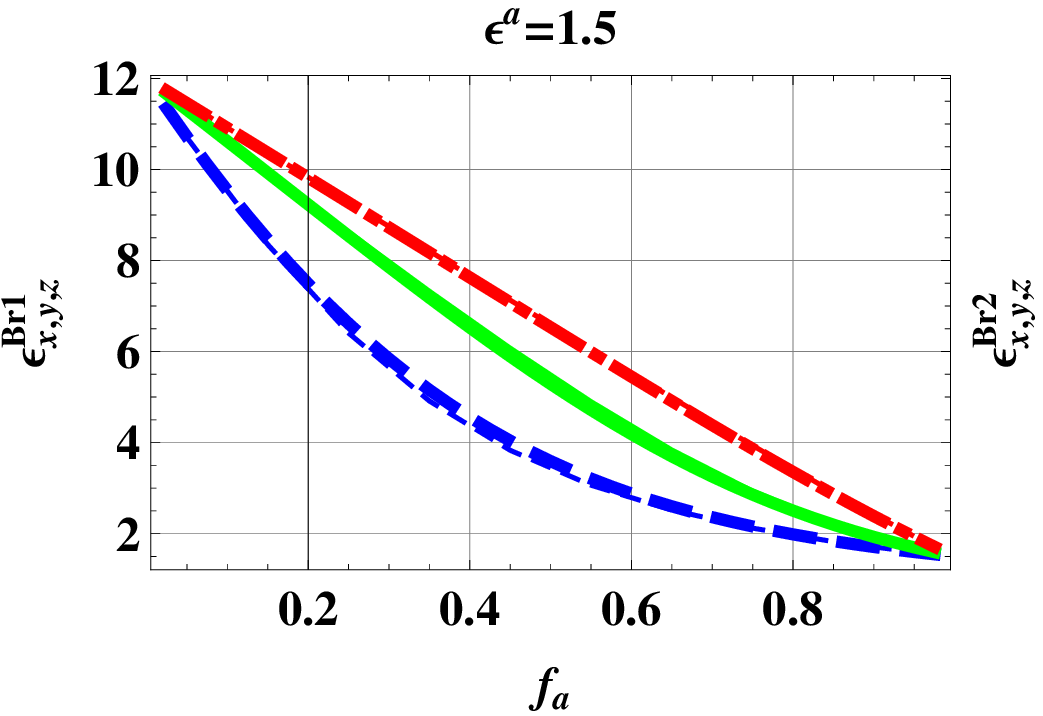} \hfill
\includegraphics[width=2.6in]{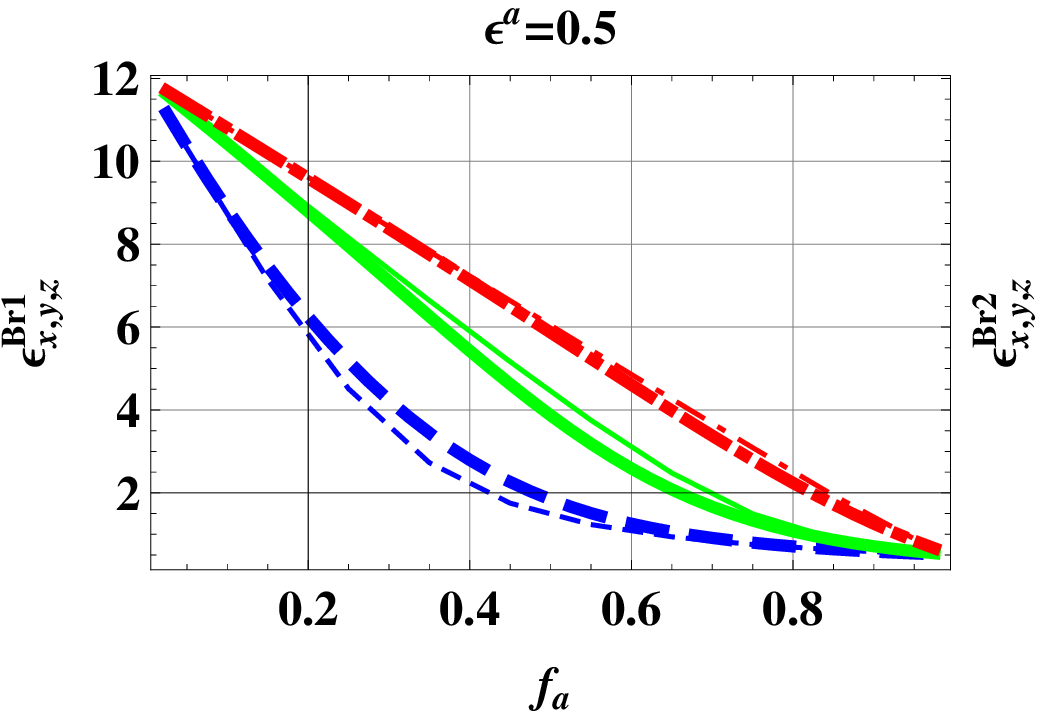}
 \caption{As Fig.~\ref{fig1} except
that $\Delta = 4.5$ and the estimates $\eps^{Br1,2}_{x,y,z}$ are
plotted versus the volume fraction $f_a \in \le 0, 1 \ri$.
 } \label{fig2}
\end{figure}

\begin{figure}[!h]
\centering
\includegraphics[width=2.6in]{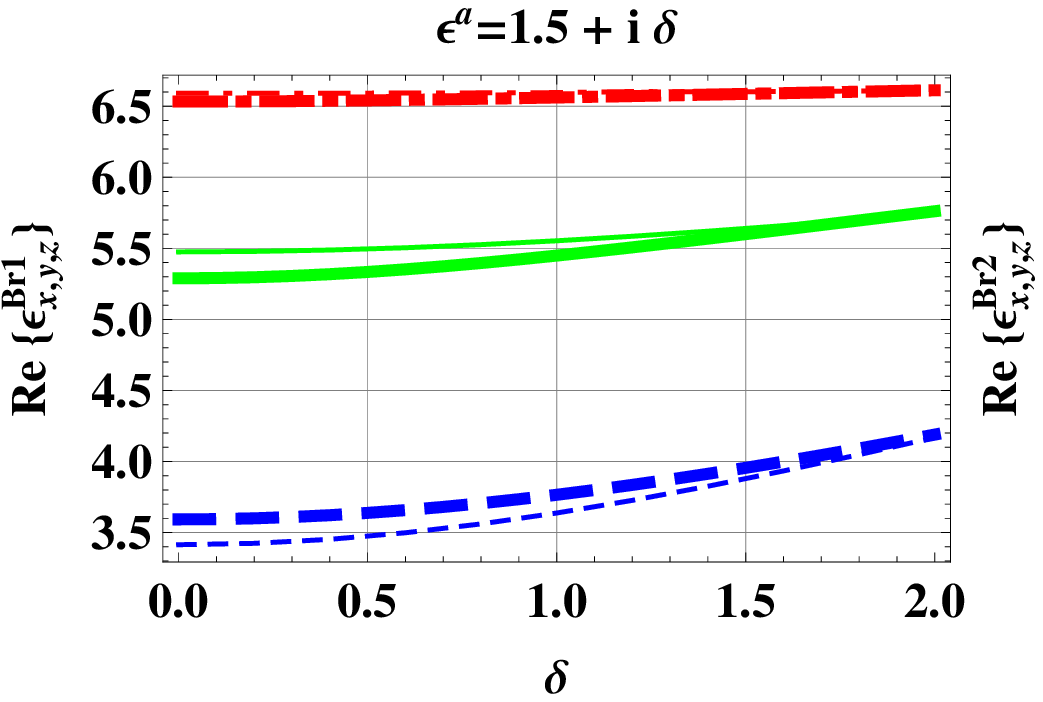} \hfill
\includegraphics[width=2.6in]{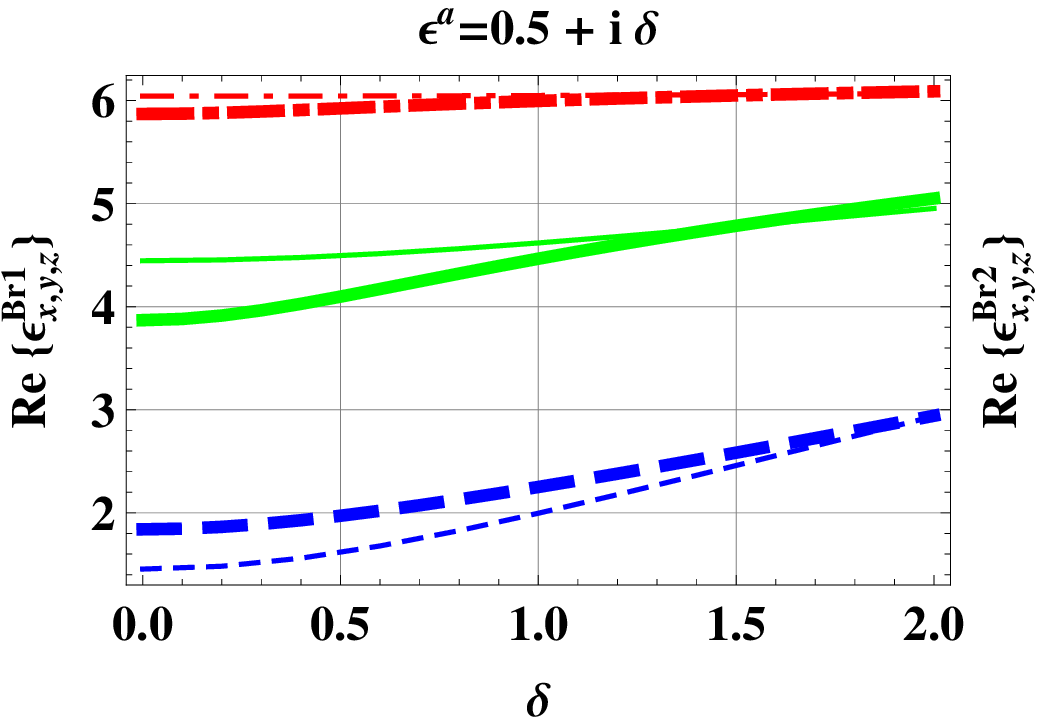} \\
\includegraphics[width=2.6in]{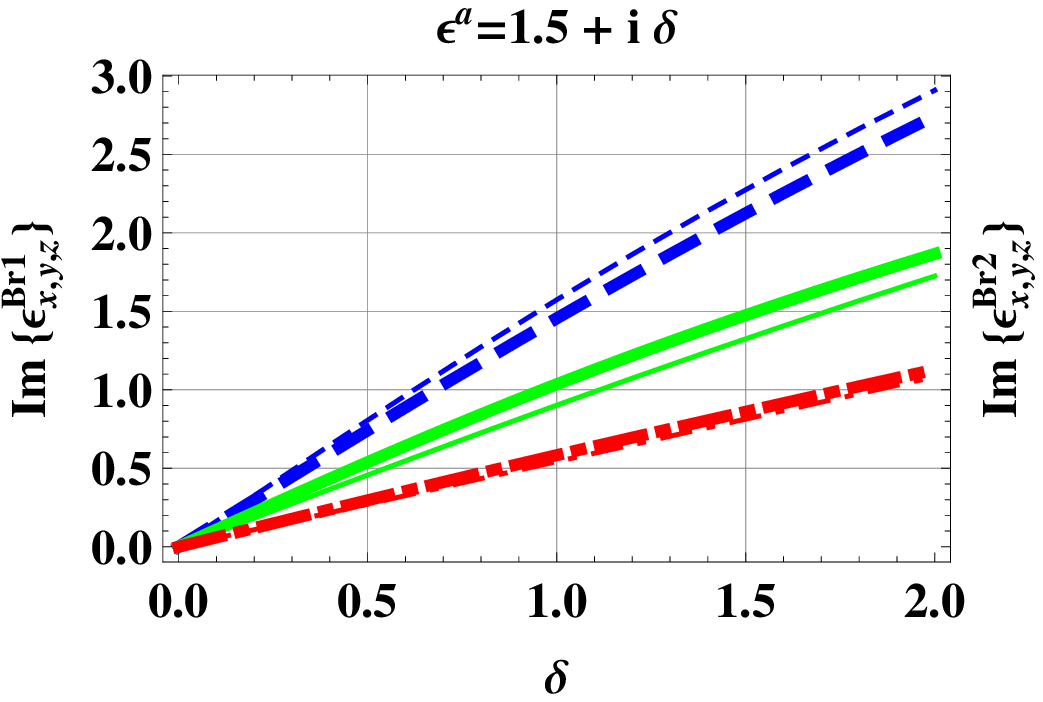} \hfill
\includegraphics[width=2.6in]{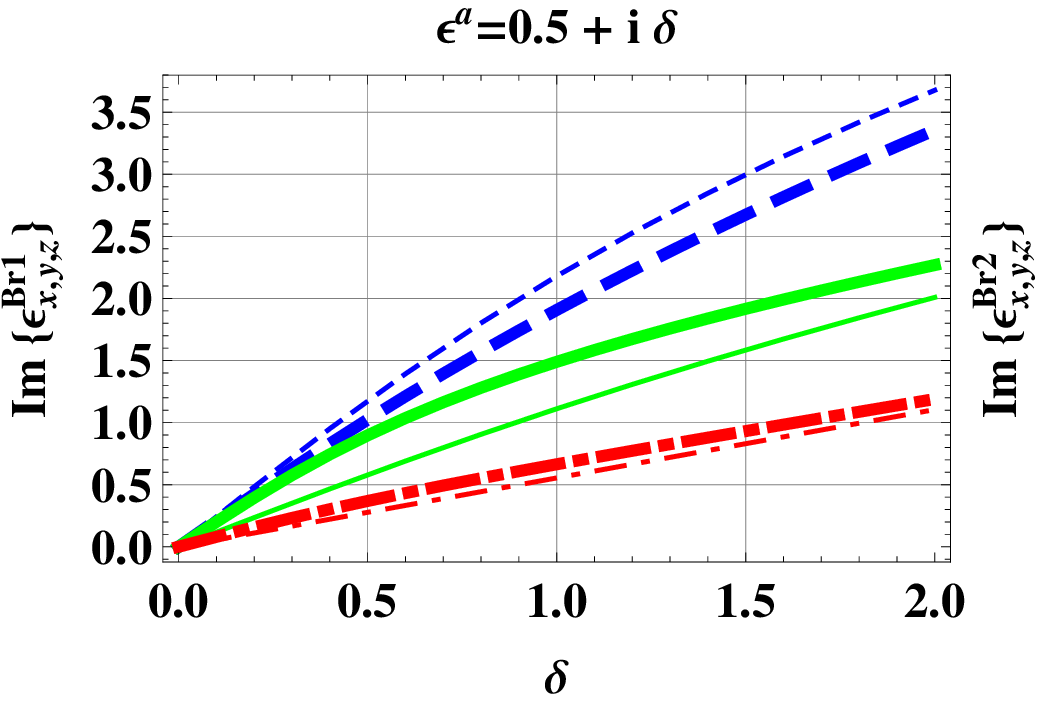} \\
 \caption{As
Fig.~\ref{fig1} except that $\Delta = 4.5$, $\eps^a \in \lec  0.5 +
i \delta, 1.5 + i \delta  \ric$, and the real and imaginary parts of
the estimates $\eps^{Br1,2}_{x,y,z}$ are plotted versus the
dissipation parameter $\delta \in \le 0, 2 \ri$. } \label{fig3}
\end{figure}

Let us illustrate the differences between the estimates
$\=\eps^{Br1}$ and $\=\eps^{Br2}$ by means of some representative
numerical results. Suppose that the shape parameters describing the
component ellipsoids have the form $U_x = 1$,  $U_y = 1 +
(\Delta/3)$, and $U_z = 2 \Delta$. Thus, the asphericity of the
ellipsoids is governed by the scalar parameter $\Delta$. We begin
with the nondissipative scenario wherein $\eps^a \in \lec 0.5, 1.5
\ric$ and $\eps^b = 12$. Also, we fix $f_a = 0.5$. Plots of the
relative permittivity parameters $\eps^{Br1,Br2}_{x,y,z}$ versus the
asphericity parameter $\Delta$ are presented in Fig.~\ref{fig1}. The
estimates $\=\eps^{Br1}$ and $\=\eps^{Br2}$ are identical for the
limiting case represented by $U_x = U_y = U_z = 1$ (i.e., for
isotropic dielectric HCMs), but differences emerge as the
asphericity of the component particles intensifies. The difference
between $\eps^{Br1}_{x}$ and $\eps^{Br2}_{x}$ grows steadily as
$\Delta$ increases, reaches a maximum for $1 < \Delta < 2$, and then
slowly shrinks as $\Delta$ increases beyond $2$. The difference
between $\eps^{Br1}_{z}$ and $\eps^{Br2}_{z}$ follows a similar
pattern. However, in the case of $\eps^{Br1}_{y}$ and
$\eps^{Br2}_{y}$, the difference increases uniformly as $\Delta$
increases. The differences between $\eps^{Br1}_{x,y,z}$ and
$\eps^{Br2}_{x,y,z}$ are generally greater for $\eps^a = 0.5$ than
for $\eps^a = 1.5$. In the former case the maximum difference is
approximately $15\%$, whereas in the latter case it is approximately
$5\%$.

We turn now to the effect of volume fraction. The calculations of
Fig.~\ref{fig1} are repeated for Fig.~\ref{fig2} except that here
the relative permittivity parameters  $\eps^{Br1,Br2}_{x,y,z}$ are
plotted versus the volume fraction $f_a$, while the asphericity parameter is
fixed at $\Delta = 4.5$. The differences between the estimates of
the two formalisms are clearly greatest at mid-range values of
$f_a$, and they are generally greater for   $\eps^a
= 0.5$ than  for $\eps^a = 1.5$.

Lastly, the effects of dissipation are considered. We repeated the
calculations of Fig.~\ref{fig1} but with $\Delta = 4.5$ and $\eps^a
\in \lec  0.5 + i \delta, 1.5 + i \delta  \ric$. Here $\delta > 0$
governs the degree of dissipation exhibited by component material
$a$. The real and imaginary parts of the relative permittivity
parameters $\eps^{Br1,Br2}_{x,y,z}$ are plotted versus the
dissipation parameter $\delta$ in Fig.~\ref{fig3}. The differences
between the real parts of the estimates $\eps^{Br1}_{x,y,z}$ and
$\eps^{Br2}_{x,y,z}$ are largest when component material $a$ is
nondissipative and they decrease uniformly as $\delta$ increases. In
contrast, the differences between the imaginary parts of the
estimates $\eps^{Br1}_{x,y,z}$ and $\eps^{Br2}_{x,y,z}$ increase as
$\delta$ increases. These differences in the imaginary parts
generally reach a maximum for mid-range values of $\delta$ and
thereafter decrease as $\delta$ increases. For both the real and
imaginary parts of the  estimates $\eps^{Br1}_{x,y,z}$ and
$\eps^{Br2}_{x,y,z}$, generally larger differences arise for $\eps^a
= 0.5 + \delta i$ than  for $\eps^a = 1.5 + \delta i$.

Thus, there are significant differences between the estimates
$\=\eps^{Br1}$ and $\=\eps^{Br2}$ when ellipsoidal component
particles are considered.  These differences may be exacerbated: (i)
if the component particles become more aspherical, (ii) at mid-range
values of the volume fractions of the component materials, and (iii)
if the HCM is dissipative. The differences between the two estimates
may be further exacerbated if one of the component materials has a
positive-valued relative permittivity which is less than unity  (or
a relative permittivity whose real part is positive-valued and less
than unity).\footnote{The parameter regime wherein one of the
component materials has a positive-valued relative permittivity
while the other has a negative-valued relative permittivity (or
likewise for the real parts of the relative permittivities in the
case of dissipative HCMs) is avoided here because the Bruggeman
formalism can deliver estimates in this regime which are not
physically plausible \c{M_Ag}.}
 Relative permittivities in
this range are associated with novel materials possessing engineered
nanostructures; these artificial materials have been the subject of
intense research lately \c{Alu,Lovat,Cia_PRB}.

The differences between the two formalisms stem from the differences
between the depolarization dyadics $\=D$ and $\=L$. The Bruggeman
formalism  conceptually employs an average--polarizability--density
approach \c{RLmotl}: Suppose the composite material has been
homogenized into an HCM. Into this HCM, let the particles of the two
component materials be dispersed in such a way as to maintain the
overally volume fractions of  $a$ and $b$. But this dispersal must
not change the effective properties of the HCM. In computing the
polarizability density dyadic of each particle, it must therefore be
assumed that the particle is surrounded by the HCM. This fact
legitimizes the use of $\=D$, which indeed contains the anisotropic
HCM's effective constitutive properties via the scalar $\rho$ of
eqn.~\r{rho}. On the other hand, use of $\=L$ to compute the
polarizability density dyadic of a particle implies that it is
surrounded by an isotropic HCM, which is clearly incorrect. Indeed,
the alternative formalism that delivers $\=\eps^{Br2}$ is an
extrapolation of the Bruggeman formalism for isotropic dielectric
HCMs \c{Smith,Granqvist}, and it lacks the rigorous basis that
underpins the estimate of $\=\eps^{Br1}$.

We have thus delineated the differences between the two formalisms
and identified one of them as the correct Bruggeman formalism. We
hope that this exposition will prevent confusion between the two
formalisms from perpetuating.

%\section{Closing remarks}

\vspace{5mm}

%\acknowledgments  AL thanks the Charles Godfrey Binder Endowment at
%Penn State for partial financial support of his research activities.

\end{document}